 \documentclass[10pt, 
 aps,amssymb,amsmath,tightenlines]{revtex4}
\usepackage{graphics}
\usepackage[pdftex]{graphicx}

\usepackage{amsfonts}
\usepackage{amssymb}

\newcommand{\ket}{\rangle}
\newcommand{\bra}{\langle}

\begin{document}

\title{Features of Quantum Mechanics on a Ring}
\author{Bernhard K. Meister}
\email{bernhard.k.meister@gmail.com}
\affiliation{ Department of Physics, Renmin University of China, Beijing, China 100872}

\date{\today }

\begin{abstract}
Aspects of quantum mechanics  on a ring are studied. Either one or two impenetrable barriers are inserted at
nodal and non-nodal points to turn the ring into either one or two infinite square wells. 
  In the process, the wave function of a particle  can change its energy, as it gets entangled with the barriers and the insertion points become nodes.
Two seemingly innocuous  assumptions representing locality and linearity are investigated. Namely, a barrier insertion at a fixed node needs no energy, and  barrier insertions can be described by linear maps.  
 It will be shown that the two assumptions are incompatible. 
\end{abstract}

\maketitle

\vspace{-1cm}

\section{Introduction}

Quantum mechanics of a particle trapped on a ring  of radius one with mass $M$ 
is described by a Hamiltonian   of the form
\begin{eqnarray}
H= - \frac{\hbar^2}{2M }\frac{d^2}{d\theta^2},\nonumber
\end{eqnarray}
defined for $\theta\in[0,2 \pi] $.
 The energy eigenvalues are $E_n=\frac{  \hbar^2}{2 M}n^2$ for $n\in \mathbb{N}$ and possess a two-fold degeneracy with the associated eigenstates
$\sin(n \theta)$ and $\cos(n \theta)$. 
Another basis combination is  $\sin(n\theta -\alpha)=\sin(n\theta) \cos(\alpha) - \cos(n\theta) \sin(\alpha)$ and $\cos(n\theta -\alpha)=\cos(n\theta) \cos(\alpha) +\sin(n\theta) \sin(\alpha)$.

The purpose of the paper is to investigate the effect of inserting impenetrable barriers at either one point, the origin, or simultaneously at both the origin and $\alpha$. 
Related investigations 
 were carried out for a barrier insertion in the one-dimensional infinite well by Bender {\it et al.}\cite{bbm} and the author \cite{bkm2011,bkm2011a}.  
 The problem of one and two barrier insertions on a ring was investigated by the author\cite{ bkm2016}.

Inserting one impenetrable barrier
turns the ring into  a one-dimensional infinite well, while  two barriers turn the ring into two separate chambers.
For a given initial state $\psi(\theta, 0)$ 
of the system, there are three possibilities for the
insertion point $\theta_0$ of the barrier. It can be a fixed or transitory node of the wave function or a general non-zero amplitude point.
By a `fixed node' we mean a point $\theta_0$ for which the time-dependent
wave function satisfies $\psi(\theta_0, t) = 0$ for all $t \geq 0$.
Wave functions that are superposition of eigenfunctions of $H$ can have zero amplitude points  that change with time called `transitory nodes'. 
%
%
We confine ourselves to `fixed nodes'  to deliver  a more striking result.
The third group of `non-nodal' points have a non-zero amplitude throughout time.\\
Cases considered in subsequent sections:\\
The first case involves one impenetrable barrier at a node. 
Since the wave
function vanishes at the insertion point, and therefore  there is no interaction with the barrier, the energy of the wave function is unchanged.  After the insertion, the configuration is turned into an one-dimensional infinite well.\\ 
Inserting a  barrier at a point that is not a node provides another example. In
this case inserting a barrier changes the energy by an amount that depends on the
rate at which the barrier is inserted. 
For conceptual simplicity, we restrict ourselves to the instantaneous insertion.\\
Finally, two barriers are inserted at a fixed node and at a non-nodal point dividing the ring into two independent infinite wells. The energy of the wave function is changed and the energy of the barriers is transformed in a complementary way to guarantee energy conservation.  The exact nature of the entanglement of the particle   with the barriers during an insertion is dependent on the insertion speed and not obvious. A variety of answers are possible. We take the different energy eigenstates of the particle in the post-insertion basis to be individually entangled with differernt barrier states. Other choices lead to other transfer energies and are experimentally distinguishable. 


What we learn from these cases is that  
 quantum mechanics on the ring cannot simultaneously satisfy two seemingly harmless  assumptions
 
1)  a barrier insertion at a  `fixed node' needs no energy, and

2)  barrier insertions can be described by linear maps,\\
 abbreviated LOC \& LIN, respectively. LOC is linked to the principle of locality. The paper provides a novel view on the clash between locality and linearity in quantum mechanics.
 Initially, we conjecture both LOC and LIN to be  correct, and then point out a contradiction, i.e. 
there is no map consistent with LIN that can account for the simultaneous insertion of two barriers (at the origin and $\alpha$) such that a barrier inserted at a nodal point requires no energy. To exploit linearity we consider both $\sin(\theta- \alpha)$ with its node at $\alpha$ as well its decomposition in the $\sin(\theta)$ and $\cos(\theta)$ basis, i.e.   $\sin(\theta-\alpha)=\sin(\theta) \cos(\alpha) - \cos(\theta) \sin(\alpha)$, whose $\sin(\theta)$ component has a node at the origin, but not at $\alpha$.  Due to linearity we can map each  component separately observing the LOC condition and sum the results. It will be shown that the combination does not coincide with the mapping of $\sin(\theta-\alpha)$ itself. 

As a consequence, one has to choose between LIN and LOC. In an earlier paper\cite{bkm2016}, we presupposed LOC and this led to the breaking of the Helstrom bound. If instead one assumes LIN, then one can distinguish between inserting one or two barriers, since simultaneous insertions can require energy even at nodal points, whereas individual insertions at fixed nodes  are energy-free.


The structure of the rest of paper is as follows. In section II   the impact of one instantaneous insertion of one barrier on a ring is studied. 
 In section III the simultaneous insertion of two barriers is considered.
  In section IV the inconsistency  of   LOC \& LIN  is proven and some implications are discussed. 
In the conclusion the result is briefly reviewed and some general comments added.

\vspace{-.5cm}
 \section{Instantaneous insertion of  one barrier }
\label{sec:1b}

In this section the insertion, considered to be instantaneous, of one barrier at both nodal and non-nodal points is reviewed. 
The nodal point insertion is dealt with first. This is easier,
since the particle wave function and energy is left unchanged -  for background material see section II of  Bender {\it et al.} \cite{bbm}, where 
 a series of results for a particle in a one dimensional box, directly applicable to quantum mechanics on a ring, were established.
  In the penultimate section, the  energy-free nature of individual insertions at fixed nodes  will be employed.



 

The situation is  more intricate  for an insertion at a non-nodal point.
Energy is needed to modify the wave function. 
 In the idealised setting considered here the required energy  is infinite. The energy localised in
the barrier point inserted at $t=0$ propagates through the system at $t > 0$
and increases the energy on the ring. The result
is a fractal wave function. The details of the calculation can be found in sections  IV \& VI of Bender {\it et al.} \cite{bbm} or in section II of \cite{bkm2011, bkm2011a}. The question how fast changes in the wave function and energy propagate through the system is of interest as one moves beyond non-relativistic quantum mechanics. 



An 
example is provided next.
The insertion at zero into the wave function $\sin(x)$, which  can also be represented as $\cos(a)\sin(x-a)+\sin(a)\cos(x-a)$, can be expanded  into the unique post insertion basis, which has lost its degeneracy due to the new boundary conditions. The expansion has the following coefficients 
\begin{eqnarray}
f_n=\frac{1}{\pi}\int_0^{2\pi} d\theta \sin(n \theta/2) (\cos(a)\sin(\theta-a)+\sin(a)\cos(\theta-a))=4\frac{ \sin(\pi n)}{n^2-4} \nonumber
\end{eqnarray}
which are zero for all $n$, except for $n=2$ with $f_2$ equal to $\pi$ and coincides with the expected result of $\sin(\theta)/\pi$.
In the coming section we move from  one to two barriers.




\vspace{-.6cm}
\section{The instantaneous insertion of two impenetrable barriers }

The case of two simultaneous insertions changing the ring into two separate infinite square wells is considered in this section. We rely in this section on the LOC assumption to restrict the transfer of energy to barriers at non-nodal points.
In the following paragraphs the wave functions  of a particle
\begin{eqnarray}
\phi(\theta)&:= &\frac{1}{\sqrt{\pi}}\sin(  \theta ),\nonumber\\
\psi(\theta)&:=&  \frac{1}{\sqrt{\pi}}\sin(\theta - \alpha),\nonumber
\end{eqnarray}  
are  evaluated before and after  the insertion of two barriers, 
 at $t=0$, 
where $\theta \in [0, 2 \pi]$, and $\alpha\in(0,\pi/2)$. 

Due to the pre-insertion symmetry  both candidate wave functions are eigenfunctions of the Hamiltonian of a particle on a ring. The symmetry is only broken by  the barriers. The barriers are inserted at the point $0$ and $\alpha$ at time $t=0$. 
The barrier inserted at point $0$   hits a fixed node of $\phi$, but the second barrier at point $\alpha$ hits a non-nodal point.
The situation is the reversed for $\psi$. 
At this stage we assume that at nodal  points no energy transfer occurs (consistent with LOC), but a barrier at a non-nodal point is associated with a change of energy.

The entanglement created during the insertion between the particle and the barriers could take different forms. 
One possible answer is to assume that the wave function in each compartment in its totality is entangled with the non-nodal barrier. Another answer, which we favour, is to have different energy eigenstates of the particle in the new basis individually entangled with the non-nodal barrier.  

An instantaneous insertion requires, due to the change of the configuration space,  an expansion of the original wave functions into the energy basis of the two separate one dimensional infinite wells. This will be carried out next.
The first expansion is in the interval $(0,\alpha)$ with the discrete energy levels $E^{\alpha}_n=\frac{  n^2 \pi^2 \hbar^2}{2 M \alpha^2}$  and the second expansion is for the interval  $(\alpha,2 \pi)$ with the discrete energy levels $E^{2\pi-\alpha}_n=\frac{  n^2 \pi^2\hbar^2}{2 M (2\pi-\alpha)^2}$ such that the first candidate wave function has directly after the insertion 
the following form 
\begin{eqnarray}
\phi_{after}(\theta):= 
\left\{
\begin{array}{lr}
\sqrt{\frac{1}{\pi}}\sum_{n=1}^{\infty}a_n \sin\Big( \frac{n \pi\theta} { \alpha}\Big)  &0< \theta < \alpha,\nonumber\\
\sqrt{\frac{1}{\pi}}\sum_{n=1}^{\infty}A_n\sin\Big( \frac{n  \pi(\theta-\alpha) }{2\pi-\alpha} \Big) & \alpha< \theta < 2\pi,\\
\end{array} 
 \right. \nonumber
\end{eqnarray}
where 
\begin{eqnarray}
a_n:=&  \frac{1}{\pi}\int_{0}^{\alpha} d\theta  \sin(\theta) \sin\Big( \frac{n \pi\theta} { \alpha}\Big)  
&= (-1)^n\ \frac{\alpha \, n }{\alpha^2-\pi^2 n^2} \sin(\alpha),\nonumber\\
A_n:=&  \frac{1}{\pi}\int_{\alpha}^{2\pi} d\theta  \sin(\theta)\sin\Big( \frac{n  \pi(\theta-\alpha) }{2\pi-\alpha} \Big) 
&= -\frac{(2\pi-\alpha)n}{(\alpha-(n+2)\pi)(\alpha+(n-2)\pi)}\sin(a).\nonumber
\end{eqnarray}
 For ease of notation we replace  $ \sin\Big( \frac{n \pi\theta} { \alpha}\Big) $ by $|\eta_n\ket$ in the interval $0< \theta < \alpha$,  and
 $\sin\Big( \frac{n  \pi(\theta-\alpha) }{2\pi-\alpha} \Big)$ by $|\chi_n\ket$ in the interval $ \alpha< \theta < 2\pi$.
The transition of $\phi$ to $|\eta_n\ket +|\chi_m\ket$ 
  is accompanied by the energy transfer to the barrier inserted at the non-nodal point $\alpha$ of
 \begin{eqnarray}
 \Delta E^{\phi}_{nm}:= 
  \frac{  \pi^2\hbar^2}{2 M }\Big( \frac{|a_n|^2}{|a_n|^2+|A_M|^2}2\frac{n^2}{\alpha^2}+ \frac{|A_M|^2}{|a_n|^2+|A_M|^2}\frac{m^2}{(2\pi- \alpha)^2}-\frac{1}{4 \pi^2}\Big).\nonumber 
 \end{eqnarray}
 For an appropriate choice of $\alpha$, e.g. $\alpha=\pi/4$,  the energy change $\Delta E^{\phi}_{nm}$ is always non-zero.

The second candidate wave function can be expanded into
\begin{eqnarray}
\psi_{after}(\theta):=
\left\{
\begin{array}{lr}
\sqrt{\frac{1}{\pi}}\sum_{n=1}^{\infty}c_n \sin\Big( \frac{n \pi \theta} { \alpha}\Big)  &0<\theta <\alpha,\nonumber\\
\sqrt{\frac{1}{\pi}}\sum_{n=1}^{\infty}C_n\sin\Big( \frac{n  \pi(\theta-\alpha) }{2\pi-\alpha} \Big) & \alpha< \theta<  2\pi,\nonumber
\end{array} 
 \right. \nonumber
\end{eqnarray}
where 
\begin{eqnarray}
c_n:=&  \frac{1}{\pi}\int_{0}^{\alpha} d\theta  \sin(\theta-\alpha) \sin\Big( \frac{n\pi\theta} { \alpha}\Big)  
&=\frac{\alpha \, n  }{\alpha^2-\pi^2 n^2}\sin(\alpha)\nonumber\\
C_n:=&  \frac{1}{\pi}\int_{\alpha}^{2\pi} d\theta  \sin(\theta-\alpha) \sin\Big( \frac{n  \pi(\theta-\alpha) }{2\pi-\alpha} \Big) 
&= (-1)^{n+1}  \frac{(2\pi-\alpha)n}{(\alpha-(n+2)\pi)(\alpha+(n-2)\pi)}\sin(\alpha).\nonumber
\end{eqnarray}
The energy transfer to $\psi$ from the barrier inserted at the  origin has the form
\begin{eqnarray}
 \Delta E^{\psi}_{nm}:= 
  \frac{  \pi^2\hbar^2}{2 M }\Big( \frac{|c_n|^2}{|c_n|^2+|C_M|^2}\frac{n^2}{\alpha^2}+\frac{|C_M|^2}{|c_n|^2+|C_M|^2} \frac{m^2}{(2\pi- \alpha)^2}-\frac{1}{4 \pi^2}\Big),\nonumber 
 \end{eqnarray}
and is identical to $\Delta E^{\phi}_{nm}$.

Due to energy conservation, there has to be a source for the energy increase of the particle. 
A laser beam could be a possible realisation for the barrier. 
The photons of the laser beam would have an energy dependent entanglement with the expanded wave function on the ring, where each combination of energy levels 
in the two chambers is linked to a complementary state for the barriers to achieve energy conservation.
The energy transfer is  to $\phi$ from the  barrier at $\alpha$ and to $\psi$ from the barrier at $0$, since each candidate wave function has its energy only modified through one specific barrier corresponding to the initial non-nodal point.

The extended wave functions including the barrier can be written before the insertion as either
\begin{eqnarray}
\Phi_{before}=  \phi\otimes \omega^{before}_0(0) \otimes  \omega^{before}_0(\alpha),\nonumber
\end{eqnarray}
or
\begin{eqnarray}
\Psi_{before}= \psi\otimes\omega^{before}_0(0) \otimes  \omega^{before}_0 (\alpha)  ,\nonumber
\end{eqnarray}
where $ \omega^{before}_0(0)$ and $\omega^{before}_0(\alpha)$ correspond to the wave functions associated with the pre-insertion 
barriers at the points $0$ and $\alpha$ respectively.
Directly after the insertion 
the  extended wave functions are transformed into
\begin{eqnarray}
\Phi_{after}=  \Big(\sum_{n=1}^{\infty} a_n |\eta_n\ket
  +\sum_{m=1}^{\infty}A_m|\chi_m\ket\Big)\otimes
\omega^{after}_{0}(0) \otimes \omega^{after}_{\Delta E_{nm}}(\alpha)\nonumber 
\end{eqnarray}
and 
\begin{eqnarray}
\Psi_{after}=\Big(  \sum_{n=1}^{\infty}c_n  |\eta_n\ket
+ \sum_{m=1}^{\infty}C_m |\chi_m\ket \Big)\otimes
\omega^{after}_{\Delta E_{nm}}(0) \otimes \omega^{after}_{0}(\alpha) ,\nonumber 
\end{eqnarray}
$\omega^{after}_0(0)$ $\&$ $\omega^{after}_0(\alpha)$ are the  barrier  wave functions, if  inserted at either a nodal point at $0$ or $\alpha$. 
$\omega^{after}_{\Delta E_{nm}}(\alpha)$ and $\omega^{after}_{\Delta E_{nm}}(0)$ correspond to barriers that transfer $\Delta E_{nm}$ to the candidate wave functions $\phi_{after}$ and $\psi_{after}$ respectively. 
The double insertion case is further studied in the next section.

\vspace{-.6cm}

\section{Inconsistency of linearity and energy-free barrier insertion at a node and its implications}
In this section, we point out the impossibility to be able to construct a certain two barrier insertion map to satisfy the following two   
  assumptions:\\ 
LOC: a barrier insertion at a 'fixed node' needs no energy, and\\
LIN:  barrier insertions can be described by  linear maps.\\
This is done by first assuming that there exist a map for which both  assumptions are true, and then pointing out a contradiction, i.e. comparing the LOC\&LIN calculation for $\sin(\theta-\alpha)$  with the  one for  $\sin(\theta) \cos(\alpha)-\cos(\theta) \sin(\alpha)$ 
and being unable to match them.

As we noticed before, $\sin(\theta)$ and $\sin(\theta-\alpha)$ have permanent nodes at zero and $\alpha$ respectively. 
 Expansion coefficients on   $(0,\alpha)$ and  $(\alpha, 2 \pi)$ for $\cos(\theta)$ are 
 \begin{eqnarray}
b_n&:=&\frac{1}{\pi}\int_{0}^{\alpha}d\theta \sin\Big(\frac{n \pi \theta}{\alpha}\Big)\cos(\theta) =\frac{\alpha n}{\alpha^2-\pi^2 n^2} \Big((-1)^n\cos(\alpha)-1\Big),
 \nonumber\\
B_n&:=&\frac{1}{\pi}\int_{\alpha}^{2\pi}d\theta \sin\Big(\frac{n \pi (\theta-\alpha)}{2\pi -\alpha}\Big)\cos(\theta) =\frac{(2\pi -\alpha)n}{(\alpha + \pi (n-2))(\pi(n+2)-\alpha)} \Big(\cos(\alpha) - (-1)^n\Big).\nonumber
\end{eqnarray}
As before, the extended wave function directly 
 after the insertions is, due to the LOC assumption, 
\begin{eqnarray}
\Psi_{after}=\Big(  \sum_{n=1}^{\infty}c_n  |\eta_n\ket
+ \sum_{m=1}^{\infty}C_m |\chi_m\ket \Big)\otimes
\omega^{after}_{\Delta E_{nm}}(0) \otimes \omega^{after}_{0}(\alpha) .
\end{eqnarray}
Replacing $\sin(\theta-\alpha)$  by $\sin(\theta) \cos(\alpha)-\cos(\theta) \sin(\alpha)$,  and assuming both LIN as well as 
LOC, one gets 
\begin{eqnarray}
\Psi_{after} 
 &=& 
\cos(\alpha)\Big( \sum_{n=1}^{\infty} a_n  |\eta_n\ket
+\sum_{m=1}^{\infty} A_m |\chi_m\ket \Big)\otimes
\omega^{after}_{0}(0) \otimes \omega^{after}_{\Delta E_{nm}}(\alpha)\nonumber
\nonumber \\
&&
-\sin(\alpha) 
 \Big(\sum_{n=1}^{\infty} b_n  |\eta_n\ket
+\sum_{m=1}^{\infty}B_m |\chi_m\ket \Big)\otimes 
\sum_{k,k'}
 R_{m,n,k,k'} 
\omega^{after}_{\Delta E_{nmk}}(0) \otimes \omega^{after}_{\Delta E_{nmk'}}(\alpha),
\end{eqnarray}
where the two barriers $\omega^{after}_{\Delta E_{nmk}}(0)$ and $\omega^{after}_{\Delta E_{nmk'}}(\alpha)$, with insertion energy transfer of $\Delta E_{nmk}(0)$ and  $\Delta E_{nmk'}(\alpha)$ respectively, split the transfer energy $\Delta E_{nm}$ between themselves, i.e. $\Delta E_{nm}=\Delta E_{nmk}(0)+\Delta E_{nmk'}(\alpha)$. For reasons of convenience,  $\alpha$ will be restricted to values with  exclusively non-zero  $\Delta E_{nm}$.  For different $k$ the $\omega^{after}_{\Delta E_{nmk}}(0)$ are mutually orthogonal. The same applies to   $\omega^{after}_{\Delta E_{nmk'}}(\alpha)$ for different $k'$.  $R_{m,n,k,k'}$ are the weights attached to different combinations of $\omega^{after}_{\Delta E_{nmk}}(0)\otimes \omega^{after}_{\Delta E_{nmk'}}(\alpha)\otimes  \Big( b_n  |\eta_n\ket+B_m |\chi_m\ket \Big)$, which  have to satisfy the formal sum $\sum_{k,k'} |R_{m,n,k,k'}|^2=1$. 

 To satisfy for this type of map both LIN \& LOC, the  total weight of each unique orthogonal product state
  for both expansions, i.e. equations (1) and (2), has to coincide. Can this be done by choosing appropriate $
R_{m,n,k,k'}$?

One  notices, $R_{m,n,k,k'}$ has for fixed $n$ \& $m$ to be zero  for all $k$ \& $k'$,  except for two  terms corresponding to energy transfers exclusively to either the barrier at $0$ or $\alpha$. First, it is the pair $\Delta E_{n,m,k}(0)=\Delta E_{nm}$ and $\Delta E_{n,m,k'}(\alpha)=0$, and second it is the pair   $\Delta E_{nmk}(0)=0$ and $\Delta E_{nmk'}(\alpha)=\Delta E_{nm}$. The corresponding weights we call  $R_{m,n}^{0}$ and $R_{m,n}^{\alpha}$  respectively.  All other $R_{m,n,k,k'}$  must be zero, since the relevant $\omega^{after}_{\Delta E_{nmk}}(0) \otimes \omega^{after}_{\Delta E_{nmk'}}(\alpha)$ 
terms, where there is a non-zero energy transfer to both barriers, only appear once in the expansion of equation (2), but not at all in equation (1). In addition,  if $n\neq n'$ and $m\neq m'$, then $( \bra \eta_n | b_n^*+ \bra \chi_m |B_{m}^* ) (b_{n'}  |\eta_{n'}\ket+B_{m'} |\chi_{m'}\ket )=0$. As a consequence,
\begin{eqnarray}
|R_{m,n}^{0}|^2+|R_{m,n}^{\alpha}|^2=1.\nonumber
\end{eqnarray}

The coefficients can be split into four sets of terms from equation (1) and (2).   The first set of terms   corresponding to $\omega^{after}_{\Delta E_{nm}}(0) \otimes \omega^{after}_{0}(\alpha)$ in the first interval, i.e. $(0,\alpha)$, is
\begin{eqnarray}
 c_n   &=&\sin(\alpha)\frac{\alpha  n}{\alpha^2-\pi^2 n^2}
\nonumber \\
&=&-\sin(\alpha)  b_n   R_{m,n}^{0}
\sin(\alpha)
=-\sin(\alpha)R_{m,n}^{0}
\frac{\alpha n}{\alpha^2-\pi^2 n^2} \Big((-1)^n\cos(\alpha)-1\Big),
\nonumber
\end{eqnarray}
which simplifies to 
\begin{eqnarray}
 1=R_{m,n}^{0} \Big(1-(-1)^n\cos(\alpha)\Big).
\end{eqnarray}
For the second set of terms corresponding to $\omega^{after}_{\Delta E_{nm}}(0) \otimes \omega^{after}_{0}(\alpha)$ in   the second interval,  i.e. $(\alpha,2\pi)$, one gets
\begin{eqnarray}
C_m & =&(-1)^{m}\frac{(2\pi-\alpha) m }{(\alpha+\pi(m-2))(\pi(m+2)-\alpha)}\sin^2(\alpha)
\nonumber\\
&=&-\sin(\alpha)   B_m R_{m,n}^{0}
=-\sin(\alpha)R_{m,n}^{0}
\frac{(2\pi -\alpha)m}{(\alpha + \pi (m-2))(\pi(m+2)-\alpha)} \Big(\cos(\alpha) - (-1)^m\Big),
\nonumber
\end{eqnarray}
to be rewritten as 
\begin{eqnarray}
 (-1)^m \sin(\alpha)=R_{m,n}^{0} \Big((-1)^m-\cos(\alpha)\Big).
\end{eqnarray}
Equation (3) and (4)  combine 
 to form 
\begin{eqnarray}
 \Big((-1)^m-\cos(\alpha)\Big)=(-1)^m \sin(\alpha)\Big(1-(-1)^n\cos(\alpha)\Big).
\end{eqnarray}
The third set of terms corresponding to $\omega^{after}_{0}(0)\otimes \omega^{after}_{\Delta E_{nm}}(\alpha) $ in the first interval leads to
\begin{eqnarray}
\cos(\alpha) a_n  &=&\cos(\alpha)(-1)^n \frac{\alpha n}{\alpha^2-\pi^2 n^2} \sin(\alpha) 
\nonumber \\
&=&\sin(\alpha) b_n   R_{m,n}^{\alpha}
=
\sin(\alpha)\frac{\alpha n}{\alpha^2-\pi^2 n^2} \Big((-1)^n\cos(\alpha)-1\Big)
R_{m,n}^{\alpha},\nonumber
\end{eqnarray}
and the fourth set of terms corresponding to $\omega^{after}_{0}(0)\otimes \omega^{after}_{\Delta E_{nm}}(\alpha) $  in the second interval is
\begin{eqnarray}
\cos(\alpha)  A_m &=&\cos(\alpha)
\frac{(2\pi -\alpha ) m}
{(\alpha + \pi (m-2))(\pi(m+2)-\alpha)} \sin(\alpha)
\nonumber\\
&=&\sin(\alpha) B_m R_{m,n}^{\alpha}
=
\sin(\alpha)
\frac{(2\pi -\alpha)m}{(\alpha + \pi (m-2))(\pi(m+2)-\alpha)} \Big(\cos(\alpha) - (-1)^m\Big)
R_{m,n}^{\alpha}.\nonumber
\end{eqnarray}
This simplifies to 
\begin{eqnarray}
(-1)^n \cos(\alpha)=R_{m,n}^{\alpha}\Big((-1)^n\cos(\alpha) - 1\Big)
\end{eqnarray}
and 
\begin{eqnarray}
\cos{\alpha}=R_{m,n}^{\alpha}\Big(\cos(\alpha) - (-1)^m\Big). 
\end{eqnarray}
By combining the equations (6) and (7) one gets
\begin{eqnarray}
(-1)^n\Big(\cos(\alpha) - (-1)^m\Big)=\Big((-1)^n\cos(\alpha) - 1\Big).
\end{eqnarray}
Equations (5) and (8) together produce 
\begin{eqnarray}
\sin(\alpha)= (-1)^n (-1)^m,
\end{eqnarray}
which cannot be satisfied simultaneously for all $n$ and $m$ for any $\alpha$ under consideration, and therefore
there is no choice of $R_{m,n,k,k'}$ satisfying the requirements.  

As a consequence, the two statements LOC \& LIN cannot both be unequivocally true for quantum mechanics on the ring with barrier insertions. 
Therefore,  one has to restrict oneself to either LOC or LIN. 
There is the possibility of both statements not to be universally true, but this will  not be explored. Each of the two statements  illuminates an aspect of quantum mechanics. This will be described in the upcoming conclusion.


\vspace{-.6cm}

\section{Conclusion}

The aim of the paper was to point out  a  paradoxical feature of quantum mechanics on a ring.
A class of linear maps   was constructed such that energy is even required for an  insertion at a fixed node, and 
implying assumptions LOC \& LIN cannot both be true. The tension between locality (LOC) and linearity (LIN) is  central to quantum mechanics, and it is brought to the fore  on a rotational symmetric configuration space, like a ring, and the respective Hamiltonian with a degeneracy. 
Experimental implementation, for example with Bose-Einstein condensate and a laser beam as a barrier, should be of interest. 

Based on the LOC condition, the barrier insertion at a fixed node is always energy-free, and implies, as discussed in an earlier paper\cite{bkm2016}, the ability to break the Helstrom bound, since pre- and post-insertion transition probabilities of the extended wave functions including the barriers were suitably modified.  
It is maybe not surprising that forfeiting linearity has such ramifications. The next paragraph, where we discard LOC, 
 but keep linearity, is maybe more startling.

If barrier insertions can be described by  linear maps (condition: LIN), then there is a difference in the energy required at the fixed node depending on the total number of simultaneous insertions. 
 To amplify the effect, we can work with a particle on a scaled up ring of radius $R$, where all the results derived for a ring of radius one still hold. We notice 
 that for appropriately chosen insertion points there is a difference in the energy needed to insert one or two simultaneously placed  barriers. One barrier at a fixed node never requires energy (section I), but two simultaneous insertions at widely separated points can even demand energy at a fixed node. This energy can be measured and, if these operations are carried out in parallel on many rings, to reduce the error probability, can be used for communication. The requirement of non-zero energy  at a fixed node forces one to conclude that another simultaneous (or earlier) barrier insertion has occurred somewhere along the ring, while  zero energy  suggests an isolated insertion. 

The paper  emphasises  the highly non-local nature of linear quantum mechanics. The energy needed to insert a barrier at a fixed node  depends in a non-local way on the amplitude of the wave function in the whole configuration space  and the actions (like other barrier insertions) undertaken at  arbitrarily separated places.  The Schr\"{o}dinger equation, a diffusion equation without an upper propagation speed, has  limitations.

Naturally, one can criticise the failure to provide a realistic time evolution; 
 only the infinitely  fast insertion was examined. In defence, one can point to the idealised nature of the proposal and the statement that more realistic examples can be viewed as an extrapolation of the procedure under consideration. The exact nature of the entanglement between the particle wave function and the barriers  is of interest and could  be checked experimentally. 
 
The following Gedankenexperiment might be instructive, because  it shows that it is possible to  construct an example were no information is transferred to the experimenter, when the potential representing the barrier is altered. 
Imagine an experimenter doing a fixed amount of work per unit of time to insert the barrier, i.e. pushes in the barrier with constant power.  
   Dependent on the test wave function the change of the potential is either larger or smaller. The potential takes on different shapes and
 affects the states in different ways without direct, energy based, leakage of information to the outside.

\hspace{-.38cm}The author wishes to express his gratitude to D.C. Brody for  stimulating discussions.

%

\end{document}